\documentclass[aps,epsfig,preprint]{revtex4-1}
\usepackage[section]{placeins}        
\usepackage{float}
 \usepackage{graphics}
 \usepackage{multirow}
\usepackage{caption}
\usepackage{epsfig}
\usepackage{pdfpages}
\usepackage{amsmath,amsthm, amssymb, latexsym}

\newcommand{\be}{\begin{equation}}
\newcommand{\ee}{\end{equation}}

\newcommand{\bea}{\begin{eqnarray}}
\newcommand{\eea}{\end{eqnarray}}

\begin{document}
\title{1D3V PIC simulation of propagation of relativistic electron beam in an inhomogeneous plasma }
 \author{Chandrashekhar Shukla$^1$}
 \email{amita@ipr.res.in}
 \author{Amita Das$^1$}
\author{Kartik Patel$^2$}
  \affiliation{$^1$Institute for Plasma Research, Bhat , Gandhinagar - 382428, India }
\affiliation{$^2$Bhabha Atomic Research Centre, Trombay, Mumbai - 400 085, India }
 
\date{\today}
\begin{abstract} 
A recent  experimental observation has shown  efficient transport of Mega Ampere of electron currents through aligned carbon nanotube arrays 
[Phys. Rev Letts. {\bf{108}}, 235005 (2012)]. The result  was subsequently interpreted on the basis of suppression of the filamentation instability in an inhomogeneous plasma [Phys. Plasmas {\bf{21}}, 012108 (2014)]. 
This inhomogeneity forms as a result of the ionization of the carbon nanotubes. In the present work a full 
  $1D3V$ Particle-in-Cell (PIC) simulations have been carried out for  the propagation of relativistic electron beams (REB) through 
an inhomogeneous background plasma. The suppression of the filamentation instability, responsible for beam divergence, is shown. 
 The simulation also confirms that in the nonlinear regime 
too the REB propagation is better when it propagates through a plasma whose density is inhomogeneous transverse to the beam. The 
role of inhomogeneity scale length, its amplitude and the transverse beam temperature etc., in the suppression of the instability is studied in detail.  

\end{abstract}
 \maketitle 

\section{Introduction}
The generation of relativistic electron beams (REB) through the interaction of a high power laser (I$\geq 10^{19}$W/$cm^{2}$) with a solid target \cite{malka,joshi} and it's 
collimated propagation through a plasma is a topic of extensive studies due to its applications in cutting edge technologies, such as 
fast ignition scheme (FIS) \cite{tabak_05} of laser fusion, compact particle acceleration \cite{mack}, fast plasma switches \cite{fuchs} and 
laser induced radiation sources \cite{park}. 

The transportation of an electron beam which carries a current much higher than the Alfven current limit $I= (mc^{3}/e)\gamma = 17{\gamma_{b}} k$ A, 
($m$ is the electron mass, $e$ is the electronic charge, $c$ is speed of light and $\gamma_{b}$ is the Lorentz factor of the beam)
is not  permitted due to its own strong magnetic field. However, in plasmas a spatially overlapping return shielding current by the background electrons 
inhibits the magnetic field generation and  permits the propagation of beams with such high forward currents. The combination of the  forward and return currents
in plasmas are, however, susceptible to several instabilities viz., two-stream \cite{bhom}, Weibel \cite{weibel}, filamentation \cite{fried} and oblique modes \cite{Fcalifano}. 
The two-stream instability is a purely longitudinal electrostatic mode which generate density stripes perpendicular to the beam flow direction. The Weibel instability is  purely 
transverse electromagnetic mode  which occurs in the system due to the temperature anisotropy.
This instability generates a magnetic field in the unmagnetized plasma system. A similar instability which generates a magnetic field in unmagnetized beam-plasma 
systems is known as the filamentation instability or beam-Weibel instability \cite{beam-weibel}. These instabilities have a detrimental influence on the propagation of the  relativistic energy
 electron beam through a plasma and hence put practical constraints. In order to suitably utilize the relativistic electron beams 
these constraints  on the propagation of beam  through a plasma needs to be overcome. A lot of experimental, theoretical and simulation work 
has been  devoted to control and suppress these  instabilities \cite{A.R.BELL,A.P.L,B.RAMA,P.MCKENNA}. 
The effect on the filamentation instability of beam density \cite{lee}, beam velocity \cite{lee}, transverse temperature \cite{temp} and collisions \cite{collision} has been widely studied. 
A recent experimental \cite{g.r.k}result, has shown many fold improvement in the propagation of a hot relativistic electron beam through an array of carbon nanotubes (CNTs).  
An explanation of this has been put forth by Mishra $\emph{et al.}$ \cite{mishra} suggesting that the inhomogeneous plasma created by the ionization of the CNTs by the front of the beam 
is responsible for the stabilization of transverse instabilities, thereby aiding the collimated propagation of the beam through longer distances compared to a homogeneous plasma.  
Here we carry out the PIC simulations to study the role of inhomogeneity with arbitrary amplitude, on the propagation of electron beam through plasma. 

 The manuscript has been organized as follows. Section II contains the details of the model configuration and governing equations. In section III, the analytical results are presented. The details of PIC simulations are 
 given in section IV. The observation of results from PIC simulation are presented in section V and it's interpretation are given in sec.VI. In last sec. VII, we conclude our results.
\section{Model Configuration and Governing Equations}
The model configuration chosen for our studies has been shown in Fig.~\ref{fig:1ddensity_profile}. The beam and the return current are chosen to flow along the $\pm \hat{y}$ respectively.
The ion density is chosen to have sinusoidal  inhomogeneity along $\hat{x}$ riding on a constant density of $n_{0}$. The amplitude of the inhomogeneity is $\epsilon$ and the spatial variations are characterized by a wavenumber $k_s = 2 \pi/\lambda_s$ as shown in  Fig.~\ref{fig:1ddensity_profile}. 
The ions are considered to be at rest at such fast electron time scales and  provide for  charge neutralization as a background species. The  density, velocity and temperature 
are denoted by $n_{\alpha}$, $v_{\alpha}$ and $T_{\alpha}$, with suffix $\alpha$ standing for  $b$ (beam)  and $p$ ( background) plasma electrons  respectively. 
   The combination of the beam - plasma system 
 can be  represented by a coupled  fluid Maxwell equations, with electrons contributing to forward and return currents treated as two distinct fluids. The normalized governing equation 
 in such a case is: 
\begin{eqnarray}
&&\frac{\partial n_{\alpha}}{\partial t}  +  \nabla \cdot\left(n_{\alpha}\vec{v_{\alpha}}\right) 
  = 0 \\
\label{con}
&&\frac{\partial \vec{p_{\alpha}}}{\partial t} + \vec{v_{\alpha}} \cdot \nabla \vec{p_{\alpha}}
  =-\left(\vec{E}+\vec{v_{\alpha}}\times \vec{B}\right) -\frac{\nabla P_{\alpha}}{n_{\alpha}} \\
 \label{mom}
&&\frac{\partial \vec{B}}{\partial t}  
  = -\nabla \times \vec{E}\\
\label{Max2}
&&\frac{\partial \vec{E}}{\partial t}  
  = \nabla \times \vec{B}- \Sigma_{\alpha}{\vec{J_{\alpha}}}
\label{Max3}
\end{eqnarray}
with
$\vec v_{\alpha}={\vec{p_{\alpha}}}/{(1+{p_{\alpha}}^{2})^{\frac{1}{2}}}$,  $\vec{J_{\alpha}}=-n_{\alpha}\vec{v_{\alpha}}$. 
The pressure $P_{\alpha}$ is provided by the equation of state. In the above equations, velocity is normalized by speed of light c, density by $n_{0}$, frequency by  $\omega_{0}=4\pi n_{0}e^2/m_{e}$ and electric and magnetic field by $E_{0}=B_{0}=m_{e}c\omega_{0}/e$ where $m_{e}$ is electron rest mass and e is electron charge.
 
In the equilibrium there is no electric and magnetic field, so there is complete charge as well as current neutralization. 
This is achieved by balancing the forward and return electron currents at each spatial location. The total charge density due to electrons is 
balanced by the background ion charge density. The background plasma has been chosen cold ($T_{0p}=0$) in all our analytical as well as in simulation studies. The profile of transverse temperature $T_{\alpha}$ in beam is chosen in such a way that gradient of pressure is zero in equilibrium. 
The temperature parallel to beam propagation direction is chosen to be zero. Thus, for an inhomogeneous beam plasma system considered by us in this work  we have the following conditions for 
equilibrium: 
 \begin{equation} 
n_{0i}(x) 
  =n_{0b}(x)+n_{0p}(x)
\label{quasi}
\end{equation}
\begin{equation} 
 \Sigma_{\alpha} n_{0 \alpha}(x)\vec{v_{0 \alpha}}=0
\label{null-current}
\end{equation}
 In equilibrium beam pressure $P_{0b}$ is chosen to be independent of x. This is achieved by choosing the beam temperature $T_{0b}(x)$ to satisfy the following condition
\begin{equation} 
P_{0b}=T_{0b}(x) n_{0b}(x)=constant=k.
\end{equation}
\begin{equation} 
 T_{0b}(x)=k/n_{0b}(x)
\label{pressure_balance}
\end{equation}
 The suffix $0$ indicates the equilibrium fields. This choice is very artificial construct, However it has been chosen to satisfy the equilibrium conditions so that comparisons with simulation can be made.   We linearize  the equations (1)-(5) to obtain linear growth rate of instability.
For the inhomogeneous case the  equation cannot be Fourier analyzed in $x$ and the eigen value can be  determined 
by solving following coupled set of differential equations: 
\begin{eqnarray}
&&\omega\gamma_{0\alpha} v_{l\alpha y}'' + \omega^3\gamma^3_{0\alpha}v_{l\alpha y}-\omega\Sigma_{\alpha}{n_{0\alpha}}v_{l\alpha y}
   +i\Sigma_{\alpha}v_{0\alpha y}(n_{0\alpha}v_{l\alpha x})' = 0 \\
\label{vy_in}
&&\eta T_{0\alpha}v_{l\alpha x}'' +
 \left(2\frac{\eta T_{0\alpha}}{n_{0\alpha}}n_{0\alpha}'+\eta T_{0\alpha}' \right){v_{l\alpha x}}'+
     \left(\frac{\eta T_{0\alpha}}{n_{0\alpha}
} n_{0\alpha}''+
\eta \frac{T_{0\alpha}'}{n_{0 \alpha}} n_{0 \alpha}'+
 \omega^2\gamma_{0\alpha}\right)v_{l\alpha x}-\Sigma_{\alpha}{n_{0 \alpha}}v_{l\alpha x}\nonumber \\
     && \hspace{2.55in}+ i\omega\gamma^3_{0\alpha}v_{0\alpha y}v_{l\alpha y}' =0 
\label{vx_in}
   \end{eqnarray}
Where $\eta$ is ratio of specific heat.
For the case of homogeneous plasma, eq.~(9) and eq.~(\ref{vx_in}) reduce to 
\begin{eqnarray}
&&\omega\gamma_{0\alpha} v_{l\alpha y}'' + \omega^3\gamma^3_{0\alpha}v_{l\alpha y}-\omega\Sigma_{\alpha}{n_{0\alpha}}v_{l\alpha y}
   +i\Sigma_{\alpha}v_{0\alpha y}n_{0\alpha}v_{l\alpha x}' = 0 \\
\label{vy_h}
&&\eta T_{0\alpha}v_{l\alpha x}'' +
 \omega^2\gamma_{0\alpha}v_{l\alpha x}-\Sigma_{\alpha}{n_{0 \alpha}}v_{l\alpha x}
    + i\omega\gamma^3_{0\alpha}v_{0\alpha y}v_{l\alpha y}' =0 
\label{vx_h}
   \end{eqnarray}
By taking Fourier transform in x, we obtain the following standard dispersion relation 
 \begin{eqnarray}
\left(\omega^{2}-k_{x}^2-\sum_{\alpha}\frac{n_{0\alpha}}{\gamma^3_{0\alpha}}\right) \left(\omega^4\gamma_{0b}\gamma_{0p}-\omega^2\gamma_{0b}\gamma_{0p}\sum_{\alpha}\frac{n_{0\alpha}}{\gamma_{0\alpha}} -\eta T_{0b}k^2_{x}( \omega^2 \gamma_{0p}-n_{0p})\right) \nonumber \\
 -k_{x}^2\left(\omega^2\gamma_{0b}\gamma_{0p}\sum_{\alpha}\frac{n_{0\alpha}v^2_{0\alpha}}{\gamma_{0\alpha}}-n_{0p}n_{0b}\sum_{\alpha}v^2_{0\alpha}+2n_{0p}n_{0b}v_{0p}v_{0b}-\eta T_{0b}k^2_{x}n_{0p}v^2_{0p}\right)=0
   \label{homogeneous-dis-rel}
 \end{eqnarray}
This equation contains two oscillatory mode and one purely growing electromagnetic mode which is known as filamentation or Weibel instability.
\section{Analytical studies}
For analytical tractability of the inhomogeneous problem a choice of sinusoidal variation riding on a homogeneous background density such as 
\begin{equation}
 n_{0i}=[1+\varepsilon cos(k_{s}x)] 
 \label{inhomo-ion}
\end{equation}
is chosen. Here 
 $\varepsilon$ is the inhomogeneity amplitude and $k_{s}=2\pi m/L_x$ (m is an integer) is the inhomogeneity wave number with $L_x$ as the system length.
 To satisfy the  quasi-neutrality condition the equilibrium beam and plasma density is chosen as 
 \begin{eqnarray}
 n_{0b} &=&\beta(1+\varepsilon cos(k_{s}x))  \nonumber \\
 n_{0p} &=&(1-\beta) (1+\varepsilon cos(k_{s}x))
 \label{inhomo-electron}
 \end{eqnarray}
 where $\beta$ is a fraction. Now choosing the perturbed fields as 
 $f_{\alpha}=\Sigma_{j}^{....,\pm2,\pm1,0}f_{\alpha j}e^{i((k+jk_{s})x-\omega t)}$ and assuming $\varepsilon$ to be small so that retaining only the first order terms (as done in the paper by 
 Mishra {\it et al.} \cite{mishra}), we can evaluate the growth rate. We evaluate the growth rate $\Gamma_{gr}$ as a function of $k$, using the same method. The plots  shown in 
 Fig.~\ref{fig:growth_rate} compare the growth rates for the homogeneous case with  inhomogeneous cases.  
Fig.~\ref{fig:growth_rate} (a) , we plot the growth rate $\Gamma_{gr}$ versus wave vector k
 for a homogeneous (solid line) and inhomogeneous cold beam plasma system for $\varepsilon$=0.1,
 for $k_{s}=\pi$(--), $2\pi$(-.-) and $3\pi$(*). The other parameters are $n_{0b}/n_{0p}=1/9$ or $n_{0b}/n_{0e}=0.1$, $v_{0b}=0.9c$, $v_{0p}=-0.1c$, $T_{0b}=0$, $T_{0p}=0$. 
 From this plot we can see that for the homogeneous case, the $\Gamma_{gr}$ increases with $k$ for small values of $k$ and saturates at $k\simeq 3$.  
 However,  for the inhomogeneous case ($\varepsilon$=0.1)  
 $\Gamma_{gr}$ for small values of $k$  is large compared to the homogeneous one. The  increase in   $\Gamma_{gr}$   with $k $ is very mild 
 and ultimately there is a  saturation at higher $k$ values. The maximum value of the growth rate  $\Gamma_{gr}$   increases with $k_{s}$ for a cold system.
 The effect of transverse beam temperature ($T_{0b\perp}=10$ keV) over the growth rate can be seen in Fig.~\ref{fig:growth_rate}(b). The growth rate $\Gamma_{gr}$ of the homogeneous hot beam and cold background plasma system (solid line) increases with $k$ for small values of $k$ but starts decreasing after $k> 0.5$ and completely stabilizes at  higher  wave numbers $(k>1)$.
 The effect of inhomogeneity on the hot beam and cold background plasma can be seen in Fig.~\ref{fig:growth_rate}(b) and Fig.~\ref{fig:growth_rate}(c). The Fig.~\ref{fig:growth_rate}(b) illustrate the effect of inhomogeneity scale length $k_s$ over the $\Gamma_{gr}$ at fixed value of  $\varepsilon$. 
 We see that for $\varepsilon=0.1$ and $ k_s=\pi$ (which is $\lambda_{s}>$c/$\omega_{p}$), the instability domain of $\Gamma_{gr}$ is much larger compared to the 
 homogeneous case. It is also observed that when the inhomogeneity scale length is further reduced to half, i.e.  $k_{s}=2\pi$(-o-) for $\varepsilon=0.1$, 
 the instability is suppressed for all values of $k$.
 In Fig.~\ref{fig:growth_rate}(c), we plot the growth rate $\Gamma_{gr}$ versus wave vector $kc/\omega_{p} $ to see the effect of $\varepsilon$ on $\Gamma_{gr}$ at $T_{0b\perp}=10$ keV.
 We see that  $\Gamma_{gr}$ reduces with increasing  $\varepsilon$.  Thus the study shows that at finite transverse beam temperature the effect of increasing $k_s$
as well as $\varepsilon$ is to stabilize the Weibel instability.  In Table - I we have tabulated the value of maximum growth rate for various cases. 
\begin{center}
{\bf{TABLE I}} \\
The maximum growth rate of filamentation instability evaluated analytically under the approximation of weak inhomogeneity amplitude. \\
\vspace{0.2in}
\begin{tabular}{c c c c c c c c c c c c c  ll}
\hline
\hline
   &$T_{0b \perp} (keV) $ \hspace{0.3in}   & $\varepsilon$  \hspace{0.3in}   &$k_{s}$ \hspace{0.3in}     &$\Gamma_{gr}$(max.)\\
 \hline
  &0.0   \hspace{0.3in}      & 0.0  \hspace{0.3in}    &   0.0   \hspace{0.3in}      &0.2006     \\
  &0.0     \hspace{0.3in}      & 0.1 \hspace{0.3in}     & $\pi$  \hspace{0.3in}      &0.2057  \\
  & 0.0    \hspace{0.3in}      & 0.1  \hspace{0.3in}    & $2\pi$  \hspace{0.3in}    &0.2074      \\
& 0.0    \hspace{0.3in}      & 0.1  \hspace{0.3in}    & $3\pi$  \hspace{0.3in}    &0.2085      \\
  &10.0  \hspace{0.3in}    &  0.0 \hspace{0.3in}    &  0.0      \hspace{0.3in}     &0.0840   \\
 &10.0 \hspace{0.3in}    & 0.1   \hspace{0.3in}   & $\pi$   \hspace{0.3in}      &0.0831  \\
 & 10.0 \hspace{0.3in}   & 0.1  \hspace{0.3in}    &  2$\pi$  \hspace{0.3in}    &  0.0770 \\ 
& 10.0    \hspace{0.3in}      & 0.1  \hspace{0.3in}    & 3$\pi$  \hspace{0.3in}    & 0.0094     \\         
& 10.0 \hspace{0.3in}   & 0.2  \hspace{0.3in}    & 3$\pi$   \hspace{0.3in}      &0.0000      \\
\hline
\end{tabular}
  \end{center} 
 It is thus  noted  that for cold beam 
 there is no significant difference between the growth rates of homogeneous and inhomogeneous cases except at long wavelengths. 
 However, when the beam temperature is chosen as finite the growth rate for the inhomogeneous 
 case significantly reduces with increasing $k_s$ as well as $\varepsilon$. We feel that the reduction in growth rate 
 is responsible for transport over long distances observed in 
 targets that had carbon nanotubes attached to them in experiments. 
 The analytical inference is further corroborated by Particle - In- Cell (PIC) simulations. The results of PIC studies are presented in the next  section.

\section{PIC simulation details} 
We have performed Particle-In-Cell simulation to study the propagation of beam in both homogeneous and inhomogeneous  plasma systems. 
The PIC code PICPSI3D used for this simulation has been developed indigenously by one of the authors (Kartik patel) \cite{patel} and has 
been used in several studies in the past \cite{Sinha}. The PICPSI3D was  generalized by us to study the propagation of electron 
beam in an inhomogeneous plasma system. Only  one dimensional variation in space (perpendicular to the beam propagation) has been chosen for 
our numerical studies here, for the purpose of comparison with recently carried out analytical linear studies.

The ions are not allowed to move in the simulations and merely provide a neutralizing background for the plasma. 
This makes the simulation faster and is a valid approximation to 
study the fast electron time scale phenomena of interest.  The uniform plasma density $n_{0}$ is chosen to be 
$10n_{c}$ where $n_{c}=1.1\times 10^{21} cm^{-3}$ is the critical density for $ 1 \mu m$ wavelength of laser light. 
The spatial simulation box length $L_x = 60$ $ c/ \omega_{0}$, where $ c/ \omega_{0}= d_{e}=5.0\times 10^{-2}\mu m$ is the skin depth 
corresponding to the density $n_{0}$.  The one dimensional simulation box is divided in $6000$ cells. The grid size is therefore equal to $ 0.01 $ $d_e$. Thus,  scales 
shorter than the  skin depth can be resolved. The total number of electrons and ions chosen for the simulations are $1800000 $ each. 
This number represents the sum of  background $n_{0p}$ and beam $n_{0b}$ electrons. The choice of inhomogeneous ion density and the separation between 
the two electron species of beam and background  are  made as per Eq.(\ref{inhomo-ion}) and Eq.({\ref{inhomo-electron})

The beam temperature $T_{b0 \perp}$ is chosen to be finite in perpendicular direction according to Eq.(\ref{pressure_balance}). The time step is decided by the Courant condition. 
The charge neutrality  as well as the null value of total current density  is ensured initially. The considered system is also field free initially as required by equilibrium configuration.
the system has an  equilibrium configuration initially.

\section{Simulation Observations }
For the homogeneous plasma density case (e.g. $\epsilon = 0.0$, $k_s = 0.0$) the system is plagued by the usual Weibel instability. 
This causes spatial separation between the forward and reverse electron currents. The separation leads to finite current density in space resulting in the growth of 
magnetic field energy. The evolution of box averaged magnetic field energy normalized by E$^2_{0}=(m_{e}c\omega_{0}/e)^2$ energy of system is shown in Fig.~\ref{fig:cold} (b), Fig.~\ref{fig:hot1} (b) and Fig.~\ref{fig:hot2} (b). 
   After an initial transient the curve settles down to a linear regime and subsequently shows saturation. 
  
The slope of the linear portion of the main curve has 
 been  employed for the evaluation of the  growth rate of the maximally unstable mode in the simulation. 
 The growth rate has been tabulated in Table - II for various cases of parameters. 
 
 \begin{center}
  {\bf{TABLE II:}} \\
   The maximum growth rate of filamentation instability calculated from PIC simulation.
\end{center}
 \begin{center}
\begin{tabular}{c c c c c c c c c c c c c c c c ll}
\hline
\hline
 &$T_{0b \perp} (keV)$   \hspace{0.3in}  &$\varepsilon$ \hspace{0.3in}   &$k_{s}$ \hspace{0.3in}  &$\Gamma_{gr}$(max.)\\
 \hline
 &0.0    \hspace{0.3in}     & 0.0   \hspace{0.3in}  &   0.0  \hspace{0.3in}     &0.2000    \\
 &0.0   \hspace{0.3in}      &0.1  \hspace{0.3in}    &$2\pi$  \hspace{0.3in}  &0.2000 \\
 &0.0  \hspace{0.3in}  &0.1  \hspace{0.3in}   &$3\pi$   \hspace{0.3in}  &0.2000 \\
&10.0   \hspace{0.3in}     &0.0  \hspace{0.3in}    & 0.0  \hspace{0.3in}    &0.0508     \\
 & 10.0\hspace{0.3in}  & 0.1  \hspace{0.3in}   &$2\pi$   \hspace{0.3in}  &0.0349     \\
&10.0\hspace{0.3in}  & 0.1  \hspace{0.3in}   &$3\pi$   \hspace{0.3in}  &0.0333     \\
&  10.0\hspace{0.3in} &  0.2 \hspace{0.3in}    &$2\pi$  \hspace{0.3in}   &0.0173    \\
&  10.0\hspace{0.3in}  &0.2  \hspace{0.3in}    &$3\pi$  \hspace{0.3in}   &0.0109 \\
\hline
\end{tabular}
  \end{center} 
   From the table as well as by following the complete evolution it is evident that  when the transverse temperature of the beam electrons is zero 
   the inhomogeneous density causes no change. 
  With  increasing value of $\epsilon$ and $k_s$ the growth rate decreases when the 
  temperature of the beam and background electrons is finite. 
  This trend is similar to the behavior of  growth rate evaluated analytically,  shown in Table - I. 
 When the plasma density is homogeneous the value of the growth rates evaluated analytically and through simulations are in good agreement. 
However, in the presence of inhomogeneity there is a small disagreement between the quantitative values.   
This can be attributed to the approximate nature of the analytical treatment, wherein  the inhomogeneity amplitude was assumed to be weak. 

  From Figures~\ref{fig:cold} (a), ~\ref{fig:hot1} (a) and ~\ref{fig:hot2} (a) it can also be seen that along with the growth of 
 magnetic field energy,  electrostatic field energy also grows. 
 The development of an electric field directed along  $x$  during the course of simulation is responsible for this electrostatic energy. 
This electrostatic field develops as a result of  the  redistribution and bunching of electron charges in 
physical $x$ space. It can be seen from the phase space plots of Figs.~\ref{fig:cold_px} and ~\ref{fig:px_hot} that  the electrons do  reorganize in physical space. 
Furthermore, the locations where these electrons get accumulated are 
the regions with maximal currents and negligible magnetic and electric field as can be seen from Fig.~\ref{fig:density_accu}

Finally we provide a comparison between the cases of homogeneous and inhomogeneous plasmas. 
It should be noted that the typical scale length of the magnetic field developed during the initial phase in the homogeneous case (Fig.~\ref{fig:magnetic} (a)) is 
of the order of the background plasma skin depth (e.g. $5.33\times10^{-2}\mu m$). For the inhomogeneous case, the scale length of the magnetic field matches initially with the inhomogeneity scale length 
defined by the choice of $k_s$ (Fig.~\ref{fig:magnetic} (b))(provided  the scale length of inhomogeneity is smaller than the skin depth) else it is determined by the typical value of skin depth. 
At later stages (the nonlinear phase of the instability), however, 
 the magnetic structures coalesce and acquire  long scales typically comparable to simulation box size  in both homogeneous (Fig.~\ref{fig:magnetic} (c)) and in-homogeneous (Fig.~\ref{fig:magnetic} (d))
 (provided the growth rate remains finite in this case) cases. 
 
 To summarize the main observations are:  (i) the inhomogeneous density causes no significant difference in the growth rate when the transverse temperature is chosen to be zero.  (ii) the  growth rate of the Weibel instability in the inhomogeneous density 
 case is reduced compared to the homogeneous case when the transverse beam temperature is finite, 
 (iii)  The momentum  $ p_x$ is typically quite large for beam electrons compared to the background plasma electrons. 
(iv) in the nonlinear regime the 
 typical profile of electrostatic field created due to electron  bunching in $x$ is similar to that of the  magnetic field. The zeros of both the fields coincide with each other 
 in space and it is these very locations where electron bunching is observed. 
 
\section{Interpretation of numerical observations}
We now provide a simplified understanding of the observations made by PIC simulations listed out in  previous section. 
 In order to understand these results we consider the 1-D limit (with only variations along $x$ being permitted) of the two fluid system of  beam and 
background electrons described in section II. In 1-D the momentum equations of the two electron species from Eqs.(1) are: 
\begin{eqnarray}
\frac{d{p}_{xb}}{dt} & = & -  \left[-\frac{\partial \phi}{\partial x}+ {v_{yb} B_z}\right] - \frac{1}{{n_b}}\frac{\partial  P_b}{\partial x } \\
\frac{d{p}_{xp}}{dt} & = & -  \left[-\frac{\partial \phi}{\partial x}+ {v_{yp} B_z} \right] \\
\frac{d{p}_{yb}}{dt} & = &- \left[ -\frac{\partial A_y}{\partial t}- {v_{xb} B_z} \right]   \\
\frac{d{p}_{yp}}{dt} &= & -  \left[ -\frac{\partial A_y}{\partial t}- {v_{xp} B_z} \right] \\
\end{eqnarray}
Here $p_{i\alpha}$ for $i = x$ and $i = y$ corresponds to the $x$ and $y$ component of momentum respectively for the beam $\alpha = b$ or plasma $\alpha = p$ 
electrons. Also $v_{i \alpha} = p_{i\alpha}/\gamma_\alpha$ (with $\gamma_\alpha$ being the relativistic factor) is the corresponding velocity. Here $P_b$ represents the transverse pressure 
which is zero for the case when the system is cold. The scalar and vector potentials are represented by $\phi$ and $\vec{A}$ respectively. In 1-D only $A_y$ 
component is finite. Thus the only finite component of magnetic field is along $\hat{z}$ and $B_z = \partial A_y/\partial x$. 

The continuity equation can be written as 
\begin{equation}
\frac{\partial n_{\alpha}}{\partial t} + \vec{v}_{x \alpha} \frac{\partial n_{\alpha}}{\partial x} + n_{\alpha} \frac{\partial v_{x \alpha}}{\partial x}= 0
\end{equation} 
The Maxwell's equation become 
\begin{eqnarray}
\frac{\partial^2 \phi}{\partial x^2} =  (\delta n_b + \delta n_p) \\
\frac{\partial^2 A_y}{\partial t^2} = \frac{\partial B_x}{\partial x} - \left[n_b v_{yb} + n_p v_{yp}\right]
\end{eqnarray}
Here $n_b$ and $n_p$ are the total densities of the beam and plasma electrons and 
$\delta n_b$ and $\delta n_p$ is the difference  between the total and   equilibrium densities  respectively. 
If one considers the transverse temperature to be zero, the linearization of the above set of equations has no other term dependent on the 
electron densities except for $E_x = - \partial \phi/\partial x$ in the momentum equation. However, Weibel being primarily an electromagnetic instability the electrostatic field is very weak. 
Thus, the predominant term in the momentum equation is due to second term of $v_{y0b} B_z$, which is not influenced by the electron density. 
Thus,  in the limit of zero temperature the homogeneous and inhomogeneous cases do not show any significant difference. 

When the transverse temperature is finite our simulations show reduction in the Weibel growth rate. In this case the pressure term in Eq.(13) 
is effective and depends on the density inhomogeneity. It has been shown by an approximate analytical studies in \cite{mishra} repeated and presented by us in Fig.2 
that the growth rate indeed decreases in the inhomogeneous case. This is the main result of our simulations which qualitatively  verifies the approximate 
results of \cite{mishra}. This can also be relevant for the observed propagation of electrons over long distances in the presence of nanowires/nanotubes in experiments\cite{g.r.k}.

As we have stated earlier along with the development of magnetic field an electrostatic field also develops. 
This happens due to the bunching of electron densities at the location of zero magnetic field as shown in Fig.~\ref{fig:density_accu}.
At the location of zero magnetic field the perturbed density shows a maxima and the electric field also passes through zero. 
This arrangement is self consistent.  The Lorentz force at these locations  vanishes  and hence a particle has a 
greater probability to accumulate over there. The location of maximum accumulation of electron density in turn results in 
the vanishing of the second derivative of electric field and for a Fourier spectrum this location should correspond to the zero of electric field. 
\section{Summary and Conclusion}
We have shown  through 1D3V PIC simulations that the growth rate of Weibel instability  gets reduced in the presence of 
density inhomogeneity. This has relevance to a  recent  experimental observation of 
  efficient transport of Mega Ampere of electron currents through aligned carbon nanotube arrays. The ionization of the carbon nanotubes by the front of laser pulse 
  produces the plasma which has inhomogeneous density. Since the Weibel instability gets suppressed in such a inhomogeneous plasma, the current 
  separation is reduced leading to the propagation of beam electrons over large distances. 
  
  This mechanism of efficient electron transport was earlier invoked by Mishra $\emph{et al.}$ \cite{mishra} wherein it was shown analytically using two fluid description 
  that the Weibel instability gets suppressed. The present work supplements it with PIC studies. Our PIC simulations support the analytical observations qualitatively. The quantitative values of the growth rate differ slightly showing 
  the approximate nature of the analysis. 
  
 {\bf{Acknowledgement:}} We thank S. Mishra for many useful discussions.
\clearpage
\newpage
 \bibliographystyle{unsrt}

\begin{figure}[1]
        \centering
                \includegraphics[width=0.65\textwidth]{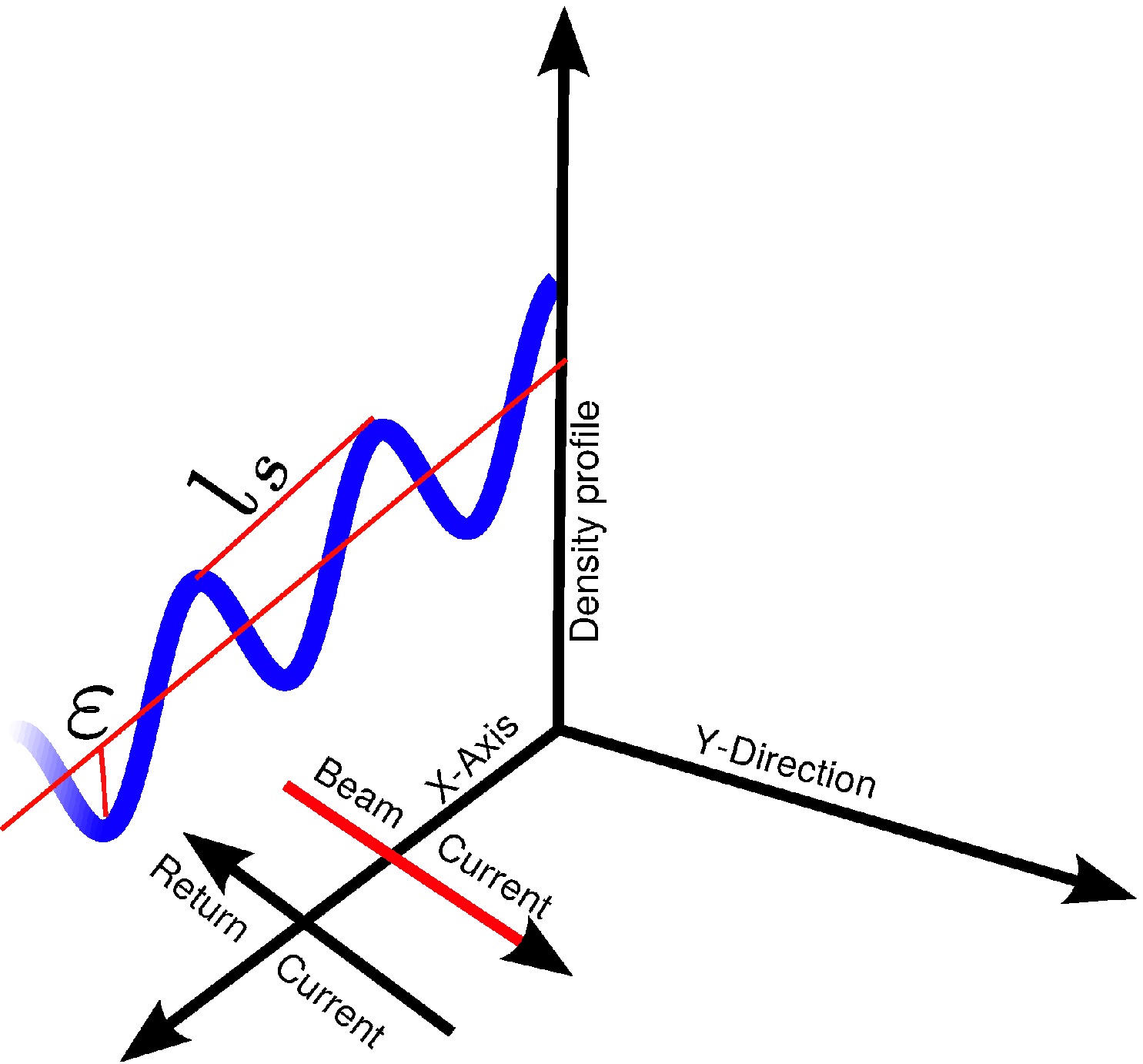} 
                 \caption {schematic of model configuration}
                 \label{fig:1ddensity_profile}
        \end{figure}%
         \begin{figure}[2]
                \includegraphics[width=\textwidth]{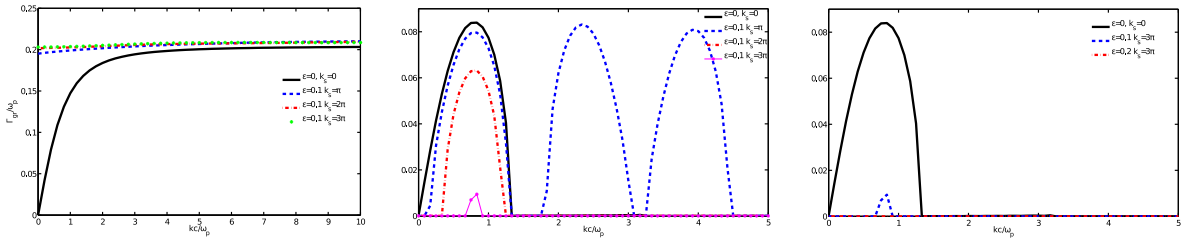} 
                  \caption{ Growth rate of the 1D filamentation instability versus the wave number k for a homogeneous and inhomogeneous beam-plasma:(a) shows growth rate for cold homogeneous beam-plasma system (solid line), for parameters $n_{0b}/n_{0e}=0.1$, $v_{0b}=0.9c$, $v_{0p}=-0.1c$ 
               and for inhomogeneous, $\varepsilon$=0.1 at $k_{s}$=$\pi$(--), $k_{s}$=$\pi$ (-.)and $k_{s}$=$\pi$(*).
                 (b) at transverse beam temperature $T_{b0\perp}$=10 keV, homogeneous(solid black), inhomogeneous for $\varepsilon$=0.1 at $k_{s}$=$\pi$(- -), $2\pi$(-.) and $3\pi$(-*)
                  (c) at transverse beam temperature $T_{b0\perp}$=10 keV, homogeneous(solid black), inhomogeneous at $k_{s}$=$3\pi$ for $\varepsilon$=0.1(- -) and 0.2 (-.)   }
                   \label{fig:growth_rate}
          \end{figure}                  
 \begin{figure}[3]
        \centering
                \includegraphics[width=\textwidth]{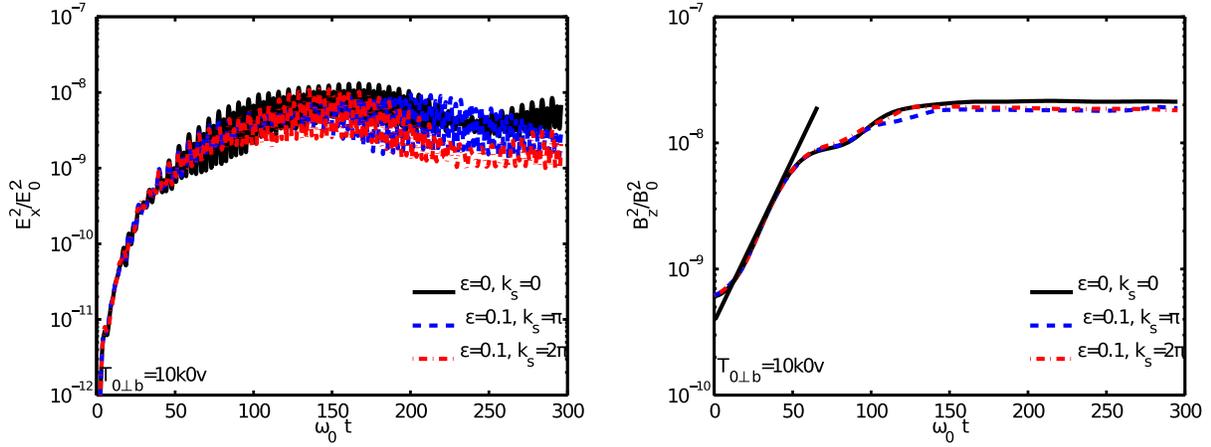} 
                 \caption { Temporal evolution of the field energy densities for cold homogeneous and inhomogeneous beam-plasma simulation. 
        (a) normalized electrostatic x-component of electric field energy 
        (b) normalized z-component of magnetic field energy     
                           }
               \label{fig:cold}
              \end{figure} 
 \begin{figure}[4]
        \centering
                \includegraphics[width=\textwidth]{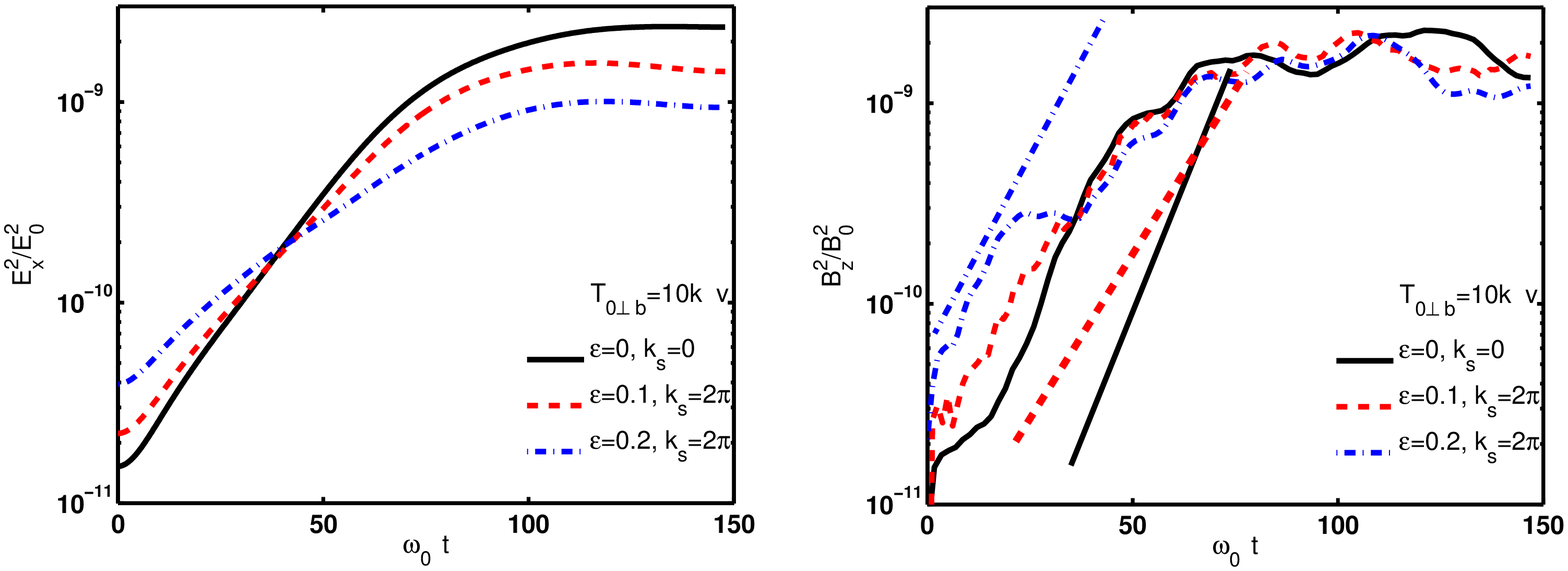} 
                 \caption { Temporal evolution of the field energy densities for hot ($T_{b0\perp}$=10 keV ) homogeneous and inhomogeneous beam-plasma simulation. 
        (a) normalized electrostatic x-component of electric field energy $E^2_{x}$ for $\varepsilon$=0, $k_{s}$=$0$ and  $\varepsilon$=0.1 and 0.2 for $k_{s}$=$2\pi$
        (b) normalized z-component of magnetic field energy $B^2_{z}$ for $\varepsilon$=0, $k_{s}$=$0$ and  $\varepsilon$=0.1 and 0.2 for $k_{s}$=$2\pi$       
                           }
               \label{fig:hot1}
              \end{figure} 
               \begin{figure}[5]
        \centering
                \includegraphics[width=\textwidth]{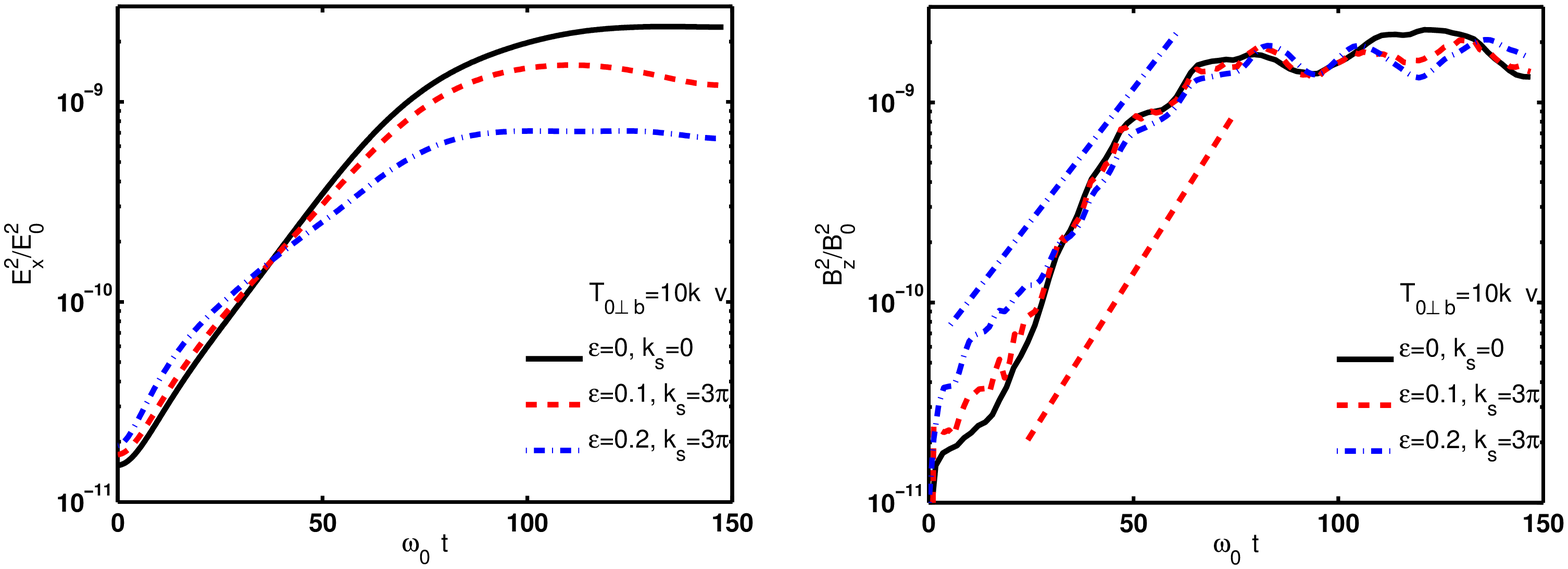} 
                \caption { Temporal evolution of the field energy densities for hot ($T_{b0\perp}$=10 keV ) homogeneous and inhomogeneous beam-plasma simulation. 
        (a) normalized electrostatic x-component of electric field energy $E^2_{x}$ for $\varepsilon$=0, $k_{s}$=$0$ and  $\varepsilon$=0.1 and 0.2 for $k_{s}$=$3\pi$
        (b) normalized z-component of magnetic field energy $B^2_{z}$ for $\varepsilon$=0, $k_{s}$=$0$ and  $\varepsilon$=0.1 and 0.2 for $k_{s}$=$3\pi$       
                           }
               \label{fig:hot2}
              \end{figure} 
             
\begin{figure}[!htb]
        \centering
                \includegraphics[width=\textwidth]{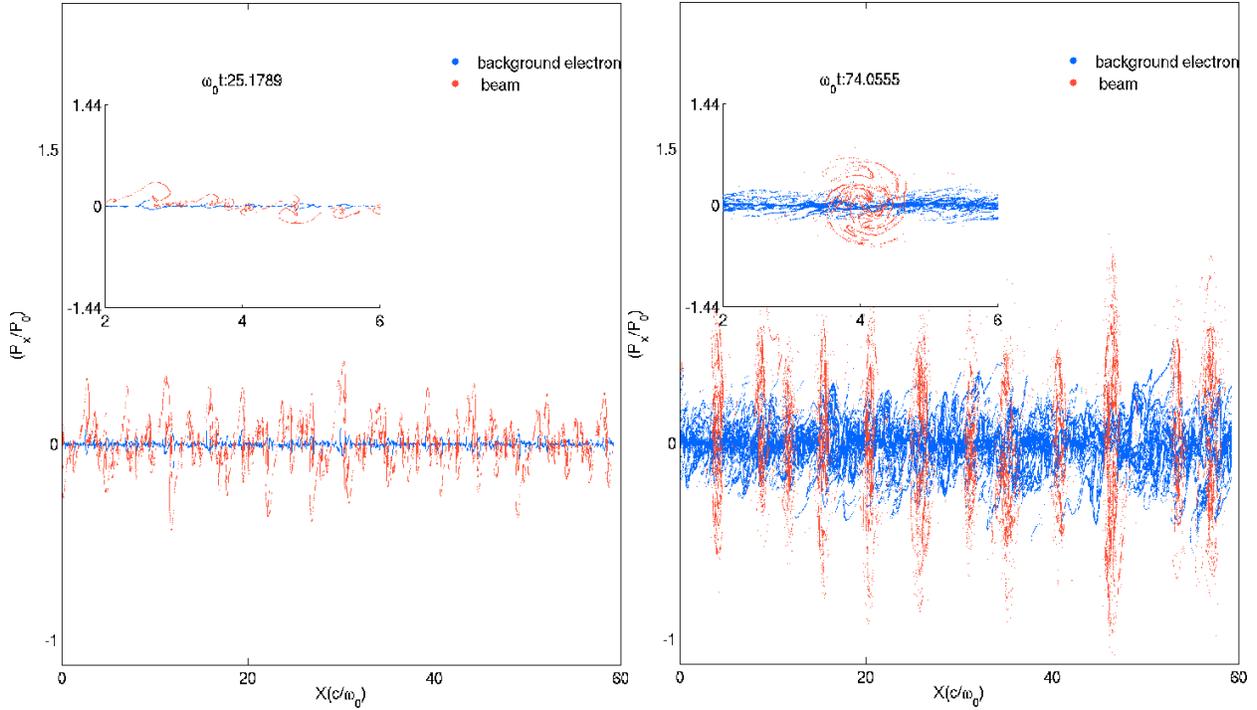} 
                 \caption { projection of f (x ,$ p_{x}$) for homogeneous cold beam plasma at $\omega_{0}t$= 25.1789 and 74.0555}
                  \label{fig:cold_px}
        \end{figure}%
        \begin{figure}[!htb]
                \includegraphics[width=\textwidth]{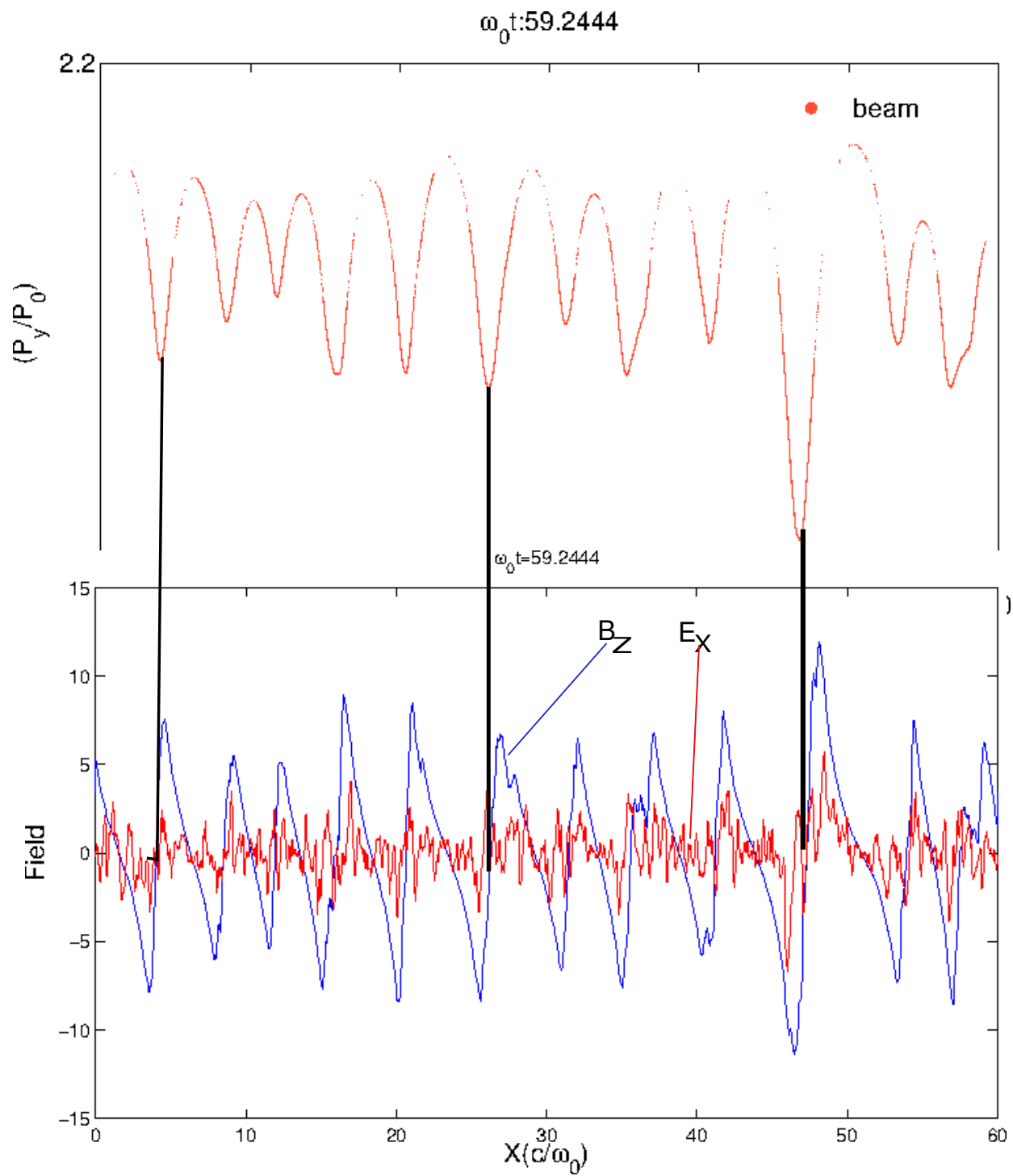}
                \caption{Bunching of perturbed electron 
                }
                 \label{fig:density_accu}
        \end{figure}
           \begin{figure}[!htb]
                \includegraphics[width=\textwidth]{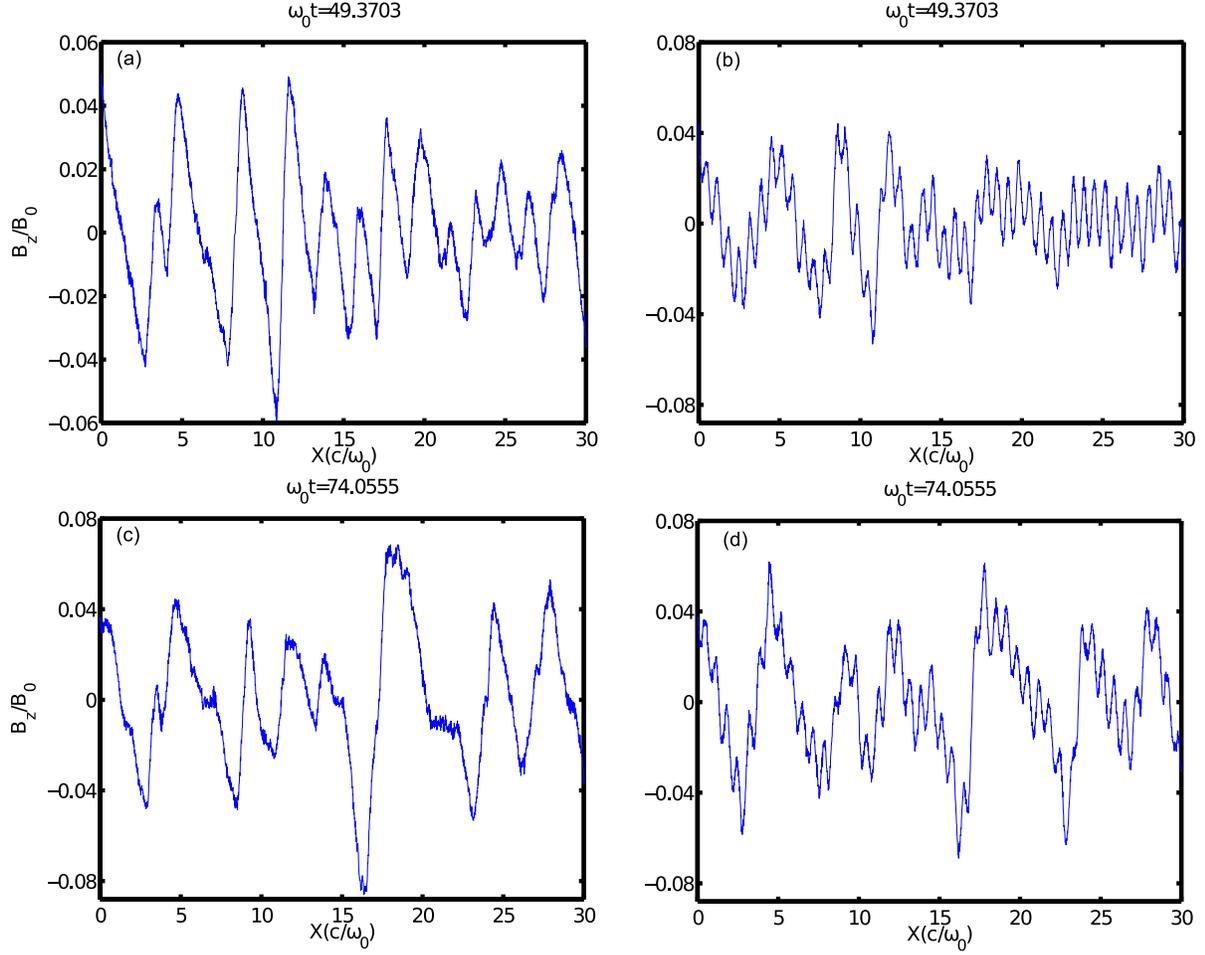}
                \caption { spatial configuration of normalized magnetic field homogeneous and inhomogeneous hot beam plasma
                (a) $\varepsilon$=0.0,  $k_{s}$=$0$  at $\omega_{0}t$=49.3703
                (b) $\varepsilon$=0.2,  $k_{s}$=$3\pi$  at $\omega_{0}t$=49.3703
                (c) $\varepsilon$=0.0,  $k_{s}$=$0$  at $\omega_{0}t$=74.0555
                (d) $\varepsilon$=0.2,  $k_{s}$=$3\pi$  at $\omega_{0}t$=74.0555
         }
           \label{fig:magnetic}        
         \end{figure}
          \begin{figure}[!htb]
                \includegraphics[width=\textwidth]{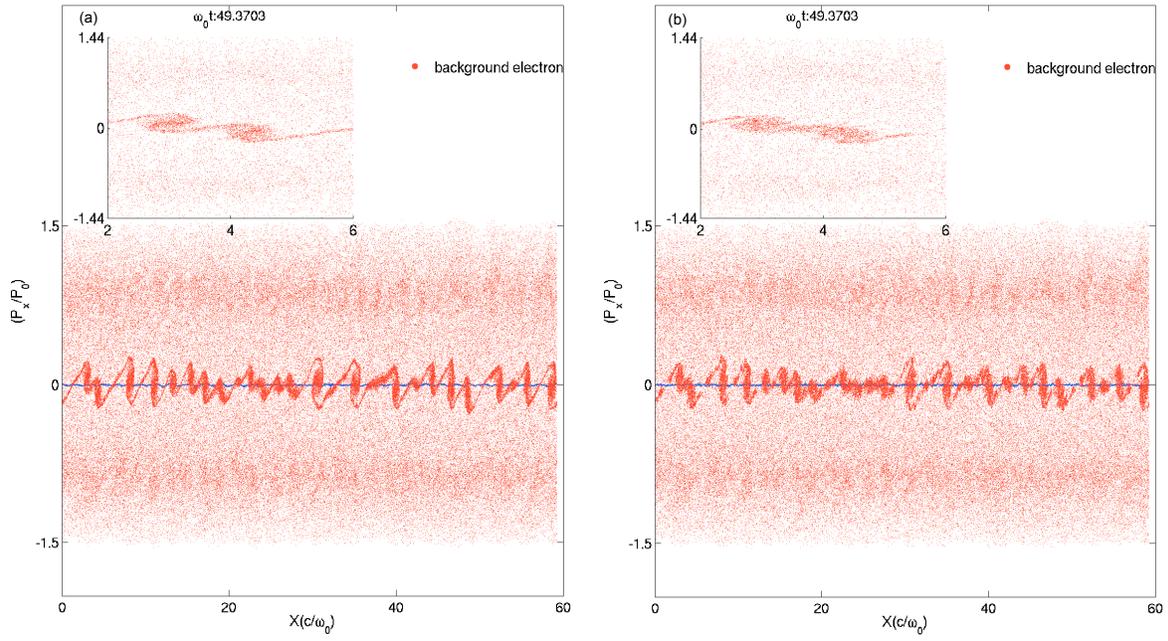} 
                \caption{projection of f (x ,$ p_{x}$) for warm system
               (a) beam with $\varepsilon$=0.0,  $k_{s}$=$0$  at $\omega_{0}t$=49.3703
                (b) beam with $\varepsilon$=0.2,  $k_{s}$=$3\pi$  at $\omega_{0}t$=49.3703
                }
                \label{fig:px_hot}  
        \end{figure}%


          
\end{document}